\documentclass[prl,twocolumn,superscriptaddress,showpacs]{revtex4}
\usepackage{graphicx}
\usepackage{amsmath}
\usepackage{amssymb}
\usepackage{dcolumn}
\usepackage{bm}
\usepackage{textcomp}
\usepackage{ulem}
\usepackage{colordvi}

\begin{document}

\title{Nuclear Charge Radius of $^{12}$Be}

\author{A. Krieger}
\affiliation{Institut f\"ur Kernchemie, Johannes Gutenberg-Universit\"at Mainz, D-55128 Mainz, Germany}

\author{K. Blaum}
\affiliation{Max-Planck-Institut f\"ur Kernphysik, D-69117 Heidelberg, Germany}

\author{M. L. Bissell}
\affiliation{Instituut voor Kern- en Stralingsfysica, KU Leuven, B-3001 Leuven, Belgium}

\author{N. Fr\"ommgen}
\affiliation{Institut f\"ur Kernchemie, Johannes Gutenberg-Universit\"at Mainz, D-55128 Mainz, Germany}

\author{Ch.~Geppert}
\affiliation{Institut f\"ur Kernchemie, Johannes Gutenberg-Universit\"at Mainz, D-55128 Mainz, Germany}
\affiliation{GSI Helmholtzzentrum f\"ur Schwerionenforschung GmbH, D-64291 Darmstadt, Germany}

\author{M.~Hammen}
\affiliation{Institut f\"ur Kernchemie, Johannes Gutenberg-Universit\"at Mainz, D-55128 Mainz, Germany}

\author{K.~Kreim}
\affiliation{Max-Planck-Institut f\"ur Kernphysik, D-69117 Heidelberg, Germany}

\author{M.~Kowalska}
\affiliation{CERN, Physics Department, CH-1211 Geneva 23, Switzerland}

\author{J.~Kr\"amer}
\affiliation{Institut f\"ur Kernchemie, Johannes Gutenberg-Universit\"at Mainz, D-55128 Mainz, Germany}

\author{T.~Neff}
\affiliation{GSI Helmholtzzentrum f\"ur Schwerionenforschung GmbH, D-64291 Darmstadt, Germany}

\author{R.~Neugart}
\affiliation{Institut f\"ur Kernchemie, Johannes Gutenberg-Universit\"at Mainz, D-55128 Mainz, Germany}

\author{G.~Neyens}
\affiliation{Instituut voor Kern- en Stralingsfysica, KU Leuven, B-3001 Leuven, Belgium}

\author{W.~N\"ortersh\"auser}
\affiliation{Institut f\"ur Kernchemie, Johannes Gutenberg-Universit\"at Mainz, D-55128 Mainz, Germany}
\affiliation{GSI Helmholtzzentrum f\"ur Schwerionenforschung GmbH, D-64291 Darmstadt, Germany}

\author{Ch.~Novotny}
\affiliation{Institut f\"ur Kernchemie, Johannes Gutenberg-Universit\"at Mainz, D-55128 Mainz, Germany}

\author{R.~S\'anchez}
\affiliation{Helmholtz-Institut Mainz, Johannes Gutenberg-Universit\"at, D-55099 Mainz, Germany}

\author{D.~T.~Yordanov}
\affiliation{CERN, Physics Department, CH-1211 Geneva 23, Switzerland}

\date{\today}

\newcommand{\fm}{\ensuremath{\textrm{fm}}}

\begin{abstract}
The nuclear charge radius of $^{12}$Be was precisely determined using the technique of collinear laser spectroscopy on the $2s_{1/2}\rightarrow 2p_{1/2,\,3/2}$ transition in the Be$^{+}$ ion. The mean square charge radius increases from $^{10}$Be to $^{12}$Be by $\delta \left\langle r_{\rm c}^2\right\rangle ^{10,12} = 0.69(5)\,\fm^{2}$ compared to $\delta \left\langle r_{\rm c}^2\right\rangle ^{10,11} = 0.49(5)\,\fm^{2}$ for the one-neutron halo isotope $^{11}$Be. Calculations in the fermionic molecular dynamics approach show a strong sensitivity of the charge radius to the structure of $^{12}$Be. The experimental charge radius is consistent with a breakdown of the $N=8$ shell closure.
\end{abstract}

\maketitle

The explanation of the magic numbers was a first remarkable success of the nuclear shell model. In stable nuclei the shell structure and therewith associated shell gaps are fairly well understood. However, in case of radioactive nuclei, far from the $\beta$-stability line, magic numbers can disappear and new magic numbers may arise. This was first noticed in case of the $N=20$ shell closure in the region of the neutron-rich Na and Mg isotopes - now called the island of inversion. Later it was found that this region is not the only island and similar observations were made at other locations within the nuclear chart (see \cite{Sor08} for a review). The magic number $N=8$ is a special case since it naturally arises in every mean-field description of nuclei. On the other hand, it is well known that light nuclei often exhibit a strongly clustered structure and mean-field calculations are not the best approach for such nuclei. Hence, a disappearance of the $N=8$ shell closure might be caused by the cluster structure in this region. Moreover, such light nuclei are in principle within the reach of \textit{ab-initio} models using realistic two- and three-body interactions. However, in the case of $^{12}$Be only more phenomenological models, {\it{e.g.}}, the shell model \cite{For11}, three-body models \cite{Nun02,Rom08}, a microscopic cluster model \cite{Duf10}, a two-center cluster model \cite{Ito08} and antisymmetrized molecular dynamics \cite{Kan03} are available, unfortunately without predictions for the charge radius.

$^{12}$Be is a key isotope in the beryllium chain, as it is located between the one-neutron halo $^{11}$Be and the two-neutron halo $^{14}$Be. $^{11}$Be has an abnormal ground state parity of $1/2^{+}$ \cite{Tal60,Gei99} indicating the breakdown of the $N=8$ shell closure with the $s_{1/2}$ halo orbit lower in energy than the $p_{1/2}$-orbit. Therefore it is expected that $sd$-shell configurations also have considerable influence in $^{12}$Be. Such an intruder-state mixing was observed in nuclear reaction experiments \cite{Nav00,Iwa00} and a dominating $sd$ configuration, {\it{i.e.}}, a quenching of the magic shell closure at $N=8$ was postulated. It was argued that the lowering of the $d_{5/2}$ level could also cause prolate deformation \cite{Iwa00} and thus a strong clustering effect inside the nucleus leading to an extended spatial structure of $^{12}$Be.
In principle, the intruding $s$-wave states can also cause halolike structures, but neither the relatively large two-neutron separation energy of $^{12}$Be ($S_{2n} = 3.67$~MeV \cite{Aud03}) nor the wide momentum distribution measured in nuclear breakup reactions \cite{Zah93} show the key characteristics of a halo structure. However, proton-scattering experiments in inverse kinematics revealed a low-density tail in the matter distribution, indicating a slight halolike character of the neutron distribution \cite{Ili11}. This is in direct contrast with the analysis of $^{11}$Be $(d,p)$ reactions, which indicated that the ground state has a small $s$-wave contribution (spectroscopic factor of $0.28^{+0.03}_{-0.07}$), whereas the long-lived excited $0_2^+$ state may exhibit an extended density tail corresponding to a much larger $s$-wave spectroscopic factor ($0.73^{+0.27}_{-0.40}$) \cite{Kan10}. In this situation additional information about the structure of $^{12}$Be is important and can be attained by a determination of the nuclear charge radius as presented in this paper.

Charge radii of light nuclei obtained from laser spectroscopy were found to exhibit clear signatures of clustering effects. This was found in the isotope chains of lithium \cite{San06}, beryllium \cite{Noe09} and neon \cite{Gei08,Mar11}, where the charge radii are in good agreement with fermionic molecular dynamics (FMD) calculations \cite{Noe11b,Zak10}. In the case of beryllium, an extension of our previous calculations for $^{7-11}$Be towards $^{12}$Be indicates that the nuclear charge radius should be very sensitive to the amount of $sd$-shell admixture into this $p$-shell nucleus. The charge radius of $^{12}$Be should be significantly smaller than for $^{11}$Be if the ground state is dominated by a $p^2$ configuration, whereas a dominance of an $(sd)^2$ configuration should lead to a larger radius. Hence, the observation of an increase in the charge radius would directly indicate a quenching of the $N=8$ shell.

We have previously reported on isotope shift measurements for the beryllium isotopes $^{7-11}$Be at ISOLDE/CERN \cite{Noe09,Zak10}. A measurement on $^{12}$Be turned out to be unfeasible at that time because of its low production rate. In this letter, we present such a measurement using an improved and more sensitive technique and we compare the extracted charge radius with new FMD calculations.

\begin{figure}[tbp]
	\includegraphics[width=\columnwidth]{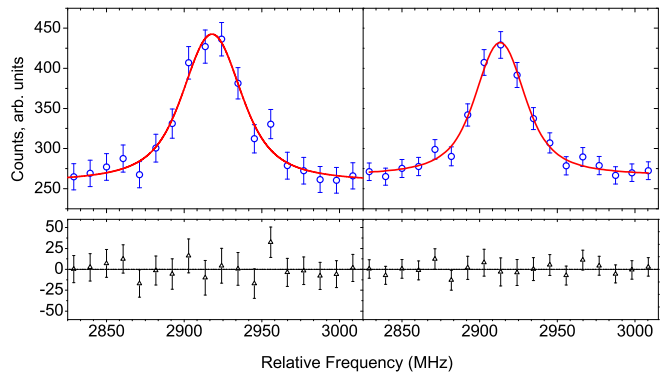}
	\caption{(Color online) Fluorescence spectra of $^{12}$Be in the D1 transition as a function of the Doppler-tuned frequency in collinear (left) and anticollinear (right) direction. A Voigt profile is fitted to the data points. The lower trace shows the residua of the fitting.}
	\label{fig:be12}
\end{figure}

We measured the optical transition frequencies of $^{12}$Be ions for the excitation $2s_{1/2}\rightarrow 2p_{1/2,\,3/2}$ (D1 and D2 lines) at 313~nm using the technique of frequency-comb based collinear and anticollinear laser spectroscopy on a Be$^{+}$ beam, which is independent of inaccuracies in calibrating the ion beam velocity \cite{Kri11}. Details of the experimental setup were presented in \cite{Noe09}. Hence, we give here a brief description, highlighting particularly the changes made for the $^{12}$Be measurement: $^{12}$Be nuclei were produced at ISOLDE/CERN by pulses of about $3 \cdot 10^{13}$ protons at 1.4~GeV impinging on a UC$_{x}$ target. On average about 1 pulse every 3-4~s was hitting the target which is coupled to the RILIS resonance ionization laser-ion source. Typically 8000 $^{12}$Be ions were extracted within 60~ms after the proton pulse, which is three times the half-life. The 40~keV ion beam was superimposed with copropagating and counter-propagating laser beams. Doppler-tuning was performed by applying a post-acceleration voltage to the fluorescence detection region (FDR). Behind the FDR, the ions were deflected and detected with a secondary electron multiplier tube (SEM). To suppress background from laser stray light, an ion-photon coincidence technique \cite{Eas86} was used. For this purpose the signals from the photomultiplier tubes were digitally delayed using a field programmable gate array (FPGA) and only pulses from the photomultiplier in coincidence with the corresponding ion signal were recorded. This avoids dead time losses from the electronic delay circuit that compensates for the time of flight of the ion from the FDR to the SEM. Only pulses from the photomultiplier in coincidence with the corresponding ion signal were recorded.\\
Typically, the resonance spectra in the D1 and D2 transitions of $^{12}$Be$^{+}$ ions were recorded with a signal to noise ({\it{S}}/{\it{N}}) ratio of about 10 after about 2 hours. In total 25 and 15 spectra for each laser beam direction were recorded for the D1 and D2 transition, respectively. Since $^{12}$Be has no hyperfine structure these spectra were fitted with a single Voigt profile as shown in Fig.~\ref{fig:be12}. The voltage scale was converted to the respective frequency scale. The full width at half maximum of a resonance is about 40~MHz, composed of a Lorentzian contribution of typically 25~MHz, which is close to the natural linewidth of about 20~MHz, and a Doppler width of 30~MHz. No asymmetry in the line shape of any isotope was observed in contrast to earlier measurements \cite{Zak10}. With the peak positions determined for both laser directions $(\nu_{a},\nu_{c})$ the transition rest frame frequency $\nu_{0}$ was calculated according to
\begin{equation}
	\nu^{2}_{0} = \nu_{a} \cdot \nu_{c}.
	\label{Eq:nu0}
\end{equation}

The absolute frequencies are then used to determine the isotope shifts $\delta\nu_{\rm IS}^{9,A} = \nu_{0}(^{A}\rm Be) - \nu_{0}(^{9}\rm Be)$. Compared to our previous work, we could considerably reduce the main source of systematic uncertainty caused by a possible photon recoil shift: The even isotopes $^{10}$Be and $^{12}$Be have no hyperfine structure and provide a closed two-level system which can repeatedly scatter photons of the laser beam. Resonant interaction with the anticollinear laser beam decelerates the ions while the interaction with photons from the collinear beam accelerates them. This shifts both frequencies, $\nu_{a}$ and $\nu_{c}$, to higher values and does not cancel when applying Eq.~(\ref{Eq:nu0}). We investigated this effect using $^{10}$Be and determined the absolute frequency for laser powers varying from 0.5 to 8~mW with a laser beam diameter of about 3~mm. No systematic shift as a function of the laser power was observed within a standard deviation of 200~kHz. With a typical laser power of 3~mW, an upper limit of 100~kHz is assumed here as a systematic uncertainty. Additionally, a possible laser-ion beam misalignment contributes with an upper limit of 500~kHz according to our previous experiments.
The isotopes $^{9,10,11}$Be were reinvestigated in the D1 line and $^{10}$Be additionally in the D2 line. The new measurements on $^{12}$Be were interspersed with measurements on $^{10}$Be throughout the beamtime, in order to check them with a well-known reference. Isotope shifts $\delta\nu_{\rm IS}^{9,A}$, as listed in Tab.~\ref{tab:results}, were calculated using the absolute frequency of $^{9}$Be as obtained as the average of all beamtimes for the respective transition. The isotope shifts determined in 2008 and 2010 agree within 300~kHz. The change in the mean square charge radius is calculated according to
\begin{equation}
	\delta \left\langle r_{\rm c}^2\right\rangle ^{9,A} = \frac{\delta\nu_{\rm IS}^{9,A}-\delta\nu_{\rm MS}^{9,A}}{F}
	\label{eq:radiusc}
\end{equation}
with the calculated mass-shift values $\delta\nu_{\rm MS}^{9,A}$, partially updated for more accurate mass measurements \cite{Rin09,Ett10}, and the electronic field-shift factor $F=-17.02$~MHz/fm$^2$, both from \cite{Yan08,Puc08}. 
While the mass-shift of $^{11}$Be includes a 208(21)~kHz contribution from nuclear polarizability \cite{Puc08}, this effect can not be calculated for $^{12}$Be for which the low-lying dipole strength distribution is not yet known. However, from preliminary breakup reaction data it is expected to be smaller than for $^{11}$Be. Hence, an additional uncertainty of 60~kHz was added to the mass-shift of $^{12}$Be.

\begin{figure}[tbp]
	\includegraphics[width=\columnwidth]{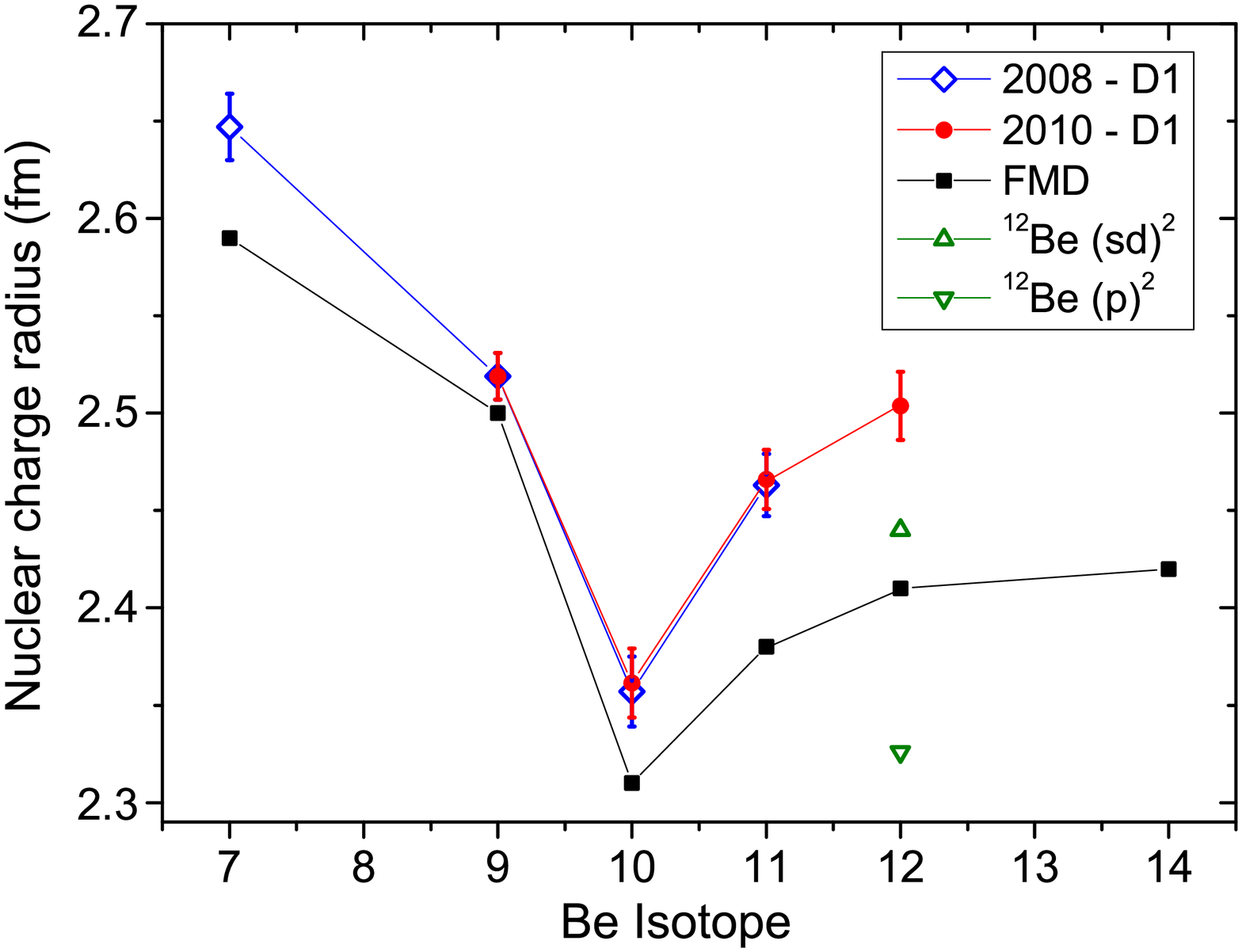}
	\caption{(Color online) Experimental charge radii of beryllium isotopes from isotope shift measurements in 2008 ({\Blue{$\diamond$}}) and in 2010 (this study \Red{$\bullet$}) compared with the FMD results (this work {\tiny \Black{$\blacksquare$}}). Charge radii calculated for the pure $p^2$-configuration (\Green{$\nabla$}) and pure $(sd)^2$ intruder configuration (\Green{$\Delta$}) are also indicated.}
	\label{fig:radii}
\end{figure}

To determine absolute rms charge radii we have used the reference radius $R_{c}(^{9} \rm Be) = 2.519(12)~{\rm fm}$ \cite{Jan72} from elastic electron scattering. Including the uncertainty of the reference radius the charge radius of $^{12}$Be was consistently extracted, from the D1 and D2 transition to be $R_{c}(^{12} \rm Be) = 2.502(16)$~fm with a relative uncertainty of less than 1~$\%$, dominated by the uncertainty of the reference radius.
\begin{table*}
	\caption{\label{tab:results}
		Changes of mean square charge radii $\delta \left\langle r_{\rm c}^2\right\rangle$ calculated from measured isotope shifts $\delta\nu_{\rm IS}^{9,A}$ relative to $^9$Be and theoretical mass-shift values $\delta\nu_{\rm MS}^{9,A}$ \cite{Puc08,Yan08,Rin09} according to Eq.~(\ref{eq:radiusc}). $ R_{\rm c}$ is the corresponding rms charge radius based on the reference radius of $^{9}$Be \cite{Jan72}. For comparison, the  matter radii $R_m$ \cite{Tan88} and the results of FMD calculations are given as well. $^{*}$ taken from \cite{Noe09}.}
	\begin{ruledtabular}
	\begin{tabular}{rcccccccccc}
		Isotope & $\delta\nu_{\rm IS}^{9,A}$ (MHz) & $\delta\nu_{\rm MS}^{9,A}$ (MHz) & $\delta \left\langle r_{\rm c}^2\right\rangle$ (fm$^2$) & $ R_{\rm c}$ (fm) & $R_c$ (FMD) (fm) & $R_m$ (Exp.)\ (fm) & $R_m$ (FMD) (fm) \\ \hline
		$^7$Be       & -49~236.88(97)$^{*}$& -49~225.779(38) & 0.66(6) & 2.646(16)  & 2.59 &
    2.31(2) & 2.37 \\[0.5ex]   
		$^9$Be       & 0             & 0               &  0        & 2.519(12) & 2.50 & 2.38(1) & 2.44 \\[0.5ex]
		$^{10}$Be    & 17~323.45(83) &  17~310.441(12)   &  -0.77(5) & 2.361(17) & 2.31 & 2.30(2) & 2.23 \\
		D2 $^{10}$Be    & ~17~325.6(16) &  17~312.569(13)   &  -0.77(9) & 2.361(24) \\[0.5ex]
		$^{11}$Be    & 31~564.71(73) &  31~560.294(24) &  -0.26(4) & 2.466(15) & 2.38 & 2.73(5) & 2.80\\[0.5ex]
		$^{12}$Be    & 43~391.46(80) &  43~390.168(39) &  -0.08(5) & 2.503(15) & 2.41 & 2.59(6) & 2.52\\
		D2 $^{12}$Be & ~43~397.0(16) &  43~395.499(39) &  -0.09(9) & 2.501(22) \\[0.5ex]
    $^{14}$Be    &                &                 &           &           & 2.42 &
    3.16(38) & 2.74
	\end{tabular}
	\end{ruledtabular}
\end{table*}
The nuclear charge radii are listed in Table \ref{tab:results} and plotted in Fig.~\ref{fig:radii}. Results of our present (\Red{$\bullet$}) and our previous work {(\Blue{$\diamond$}}) are compared with new calculations in the FMD approach.

To understand the consequences of the result for the structure of $^{12}$Be, we reinvestigated our FMD calculations for the beryllium isotopic chain. FMD \cite{Nef08} uses a Gaussian wave-packet basis for the single-particle states that allows to describe nuclei with clustering and halos in a consistent picture. Good quantum numbers are obtained by projecting the intrinsic many-body basis states on parity, angular and total linear momentum. The parameters of the basis states are obtained by variation after projection for the spins of the low-lying states for each isotope. Additional basis states are obtained by constraining the quadrupole moment of the intrinsic states. The full wave functions are obtained in a multiconfiguration mixing calculation with 30-40 basis states. 

In \cite{Zak10} we performed FMD calculations that were able to describe the evolution of the charge radii in the beryllium isotopes up to $^{11}$Be. However with the effective interaction employed there, the gap between $p$- and $sd$-shell-orbits is too large and the parity inversion in $^{11}$Be could not be described. As the competition between $p$- and $sd$-configurations plays an essential role in $^{12}$Be, a different interaction is used in this work. It is derived from the Argonne~V18 potential \cite{Wir95} in the unitary correlation operator method \cite{Nef03} with correlation functions obtained from a Hamiltonian evolved in the similarity renormalization group \cite{Rot10}. This UCOM(SRG) interaction has been used successfully in FMD for a calculation of the $^3$He($\alpha$,$\gamma$)$^7$Be capture reaction \cite{Nef11}. However no-core shell model calculations \cite{Rot10} show that the effective spin-orbit splittings with this realistic two-body interaction are too small. To correct for that deficiency without employing three-body forces we introduce a phenomenological parameter $\eta$ to change the strength of the spin-orbit force in the $S=1$, $T=1$ channel. A factor $\eta \approx 2$ works remarkably well for the beryllium isotopes. This modification of the spin-orbit force leads to increased binding and it shifts the relative position of $p$- and $sd$-orbits. The best agreement with experimental separation energies is obtained for a strength factor which changes from $\eta = 1.9$ (for $^9$Be) to $\eta = 2.2$ (for $^{12}$Be). With the exception of $^{12}$Be properties like radii and transitions depend only very weakly on the details of the spin-orbit force. The $A$-dependence of $\eta$ indicates that this procedure might indeed simulate the role of three-body forces. 

\begin{figure}
	\includegraphics[width=\columnwidth]{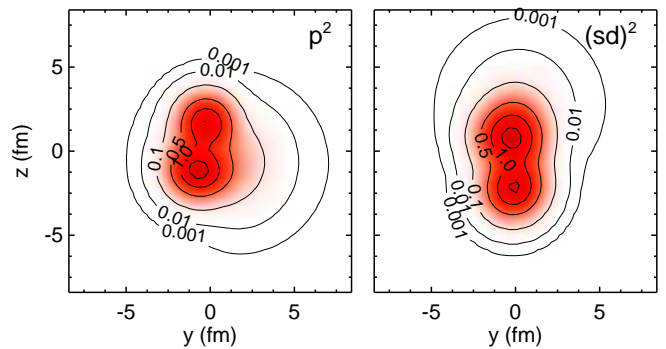}
	\caption{Cuts through the intrinsic densities of the $^{12}$Be states in $p^2$ (left) and $(sd)^2$ configurations (right) obtained by variation after projection. The contour lines are in units of nuclear matter density.}
	\label{fig:intrinsic}
\end{figure}

The intrinsic states of the beryllium isotopes show a more or less pronounced $\alpha$-cluster structure as already discussed in \cite{Zak10}. Depending on the isotope the distance between the $\alpha$-clusters is different. This effect is the reason for the large decrease in the charge radius from $^9$Be to $^{10}$Be. The jump between $^{10}$Be and $^{11}$Be on the other hand, is mainly explained by the motion of the $^{10}$Be core with respect to the total center-of-mass of the core and the halo neutron in an extended $s$-orbit. However, we also find a small contribution related to core excitation from the $d$-wave component. In the case of the $0^+$ states of $^{12}$Be we find in the FMD calculation two different local minima that correspond essentially to a $p^2$ and an intruder $(sd)^2$ configuration as shown in Fig.~\ref{fig:intrinsic}.  Whereas the charge radius of the $p^2$ configuration with 2.33~\fm~is only slightly larger than in $^{10}$Be, the $(sd)^2$ configuration has a significantly larger charge radius of 2.44~\fm, which is caused by both an increased distance of the $\alpha$-clusters and a correlation with the neutrons which leads to an $\alpha$-$^8$He like structure. The $(sd)^2$ configuration in $^{12}$Be is very different from the picture in $^{11}$Be where a very small neutron separation energy favors the appearance of an $s$-wave halo. In $^{12}$Be the two neutrons are much better bound and the FMD single-particle orbits are linear combinations of $d$- and $s$-orbits. This is also reflected in the matter radius that is significantly smaller in the $^{12}$Be $(sd)^2$ configuration than in $^{11}$Be. As the $p^2$ and $(sd)^2$ configurations are very close in energy, the mixing is strong and the charge radius of the ground state depends sensitively on the mixing.
This is also the case for other observables, like the monopole matrix element between the two $0^+$ states and the quadrupole transition strengths of the two $0^+$ states to the first $2^+$ state. With the strength of spin-orbit force adjusted to the separation energy we obtain a good agreement for all transitions as shown in Table~\ref{tab:be12properties}. The $(sd)^2$ configuration contributes about 70\% to the ground state. This is reflected in the charge radius which is larger than the calculated $^{11}$Be charge radius by about 0.03~\fm. The FMD calculations are therefore able to describe the experimental trend of the charge radii from $^9$Be to $^{12}$Be. However, the absolute values are somewhat too small, especially for the heavier isotopes. This might be related to deficiencies of the effective two-body interaction which is known to saturate at too high densities for heavier nuclei \cite{Rot10}.

\begin{table}[b]
 \begin{ruledtabular}
 \caption{Two-neutron separation energy and $M(E0)$ and $B(E2)$ transition strengths in $^{12}$Be.}
	\label{tab:be12properties}
	\begin{tabular}{lcl}
		& FMD & Exp \\ \hline
              $S_{2n} (\mathrm{MeV})$ & 3.81 & 3.673
							\cite{Aud03}\\
		          $B(E2; 2_1^+ \rightarrow 0_1^+) (e^2\,\fm^4) $ & 8.75 & 8.0(3.0) \cite{Ima09}\\
	            $B(E2; 0_2^+ \rightarrow 2_1^+) (e^2\,\fm^4) $ & 7.45 & 7.0(6) \cite{Shi07}\\
	            $M(E0; 0_1^+ \rightarrow 0_2^+) (e\,\fm^2) $ & 0.90 & 0.87(3) \cite{Shi07}\\
	\end{tabular}
	\end{ruledtabular}
	
\end{table}

While the charge radius increases from $^{11}$Be to $^{12}$Be, the matter radius decreases considerably \cite{Tan88}, which is also in accordance with our FMD calculations.

In summary, we investigated the $2s_{1/2}\rightarrow 2p_{1/2,\,3/2}$ transitions in $^{12}$Be ions using combined collinear and anticollinear laser spectroscopy to extract the isotope shifts of $^{12}$Be relative to $^{9}$Be. By combining the experimental data with recent mass-shift calculations we obtained the change in the mean square charge radius from $^{10}$Be to $^{12}$Be of $0.69(5) \fm^2$ resulting in an absolute radius of $2.502(16) \fm$. This increase of the charge radius provides important new information for understanding the structure of $^{12}$Be. In FMD model calculations the increase of the charge radius is related to a breakdown of the $N=8$ shell closure with an $(sd)^2$ admixture of about 70\%. This is also consistent with other observables such as the lifetime of the isomeric $0^+$ state and the quadrupole transition strengths to the first $2^+$ state.

\begin{acknowledgments}
This work was supported by the Helmholtz Association (VH-NG148), the German Ministry for Science and Education  (BMBF, 06MZ9178I), the European Union Seventh Framework (contract no. 262010) and the BriX IAP Research Program No. P6/23 (Belgium). A.~Krieger acknowledges support from the Carl-Zeiss-Stiftung (AZ:21-0563-2.8/197/1). We are indebted to G.W.F.~Drake and K.~Pachucki for providing the latest mass-shift calculations and we thank the ISOLDE technical group, Menlo Systems and Sirah Laser for their support.
\end{acknowledgments}

\end{document}